\documentclass[a4paper, 12pt]{article}

\makeatletter
\def\Hv@scale{.9}  % Skalierung für Helvetia, da diese in Verbindung mit Times zu groß wirkt
\makeatother
\makeatletter
\def\section{\@startsection{section}{1}{\z@}%
  {-2ex \@plus -1ex \@minus -.2ex}%
  {1ex \@plus.2ex}%
  {\raggedright\Large\bfseries\sffamily}}
\def\subsection{\@startsection{subsection}{2}{\z@}%
  {-3ex\@plus -1ex \@minus -.2ex}%
  {.8ex \@plus .2ex}%
  {\raggedright\normalsize\bfseries\sffamily}}
\let\subsubsection\undefined
\def\paragraph{\@startsection{paragraph}{4}{\z@}%
  {.5ex \@plus1ex \@minus.2ex}%
  {-.5em}%
  {\raggedright\normalsize\bfseries\sffamily}}
\makeatother

\usepackage{amsmath,amsthm}

\newtheoremstyle{plain}
  {0.3\topsep}   % ABOVESPACE, 0.5\topsep für remark
  {0.3\topsep}   % BELOWSPACE, 0.5\topsep für remark
  {\itshape}  % BODYFONT, \normalfont für remark, def.
  {}       % INDENT (empty value is the same as 0pt)
  {\sffamily\bfseries} % HEADFONT, \itshape für remark
  {.}         % HEADPUNCT
  {5pt plus 1pt minus 1pt} % HEADSPACE
  {}          % CUSTOM-HEAD-SPEC

\theoremstyle{plain}
\newtheorem{thm}{Theorem}[section]
\newtheorem{lem}[thm]{Lemma}
\newtheorem{cor}[thm]{Corollary}
\newtheorem{asm}[thm]{Assumption}
\newtheorem{dfn}[thm]{Definition}

\newenvironment{prf}[1][Proof]{\vspace{-.75\topsep}\begin{proof}[\bfseries\sffamily\upshape #1]}{\end{proof}\vspace{-.5\topsep}}

\def\(#1\){\begin{align*}#1\end{align*}}
\newcommand\eqlabel[1]{\stepcounter{equation}\tag{\theequation}\label{#1}}

\AtBeginDocument{
  \setlength\floatsep{6pt}
  \setlength\textfloatsep{6pt}
  \setlength\intextsep{6pt}
  
  \setlength\abovedisplayskip{4pt plus 2pt}
  \setlength\belowdisplayskip{4pt plus 2pt}
  \setlength\abovedisplayshortskip{4pt plus 2pt}
  \setlength\belowdisplayshortskip{4pt plus 2pt}    
  
  \newlength\oldfontdimen
  \setlength\oldfontdimen{\fontdimen2\font}
  \fontdimen2\font=.75\oldfontdimen
  \let\oldeverypar\everypar
  \newtoks\everypar
  \oldeverypar{\the\everypar\looseness=-1\relax}
  \renewenvironment{thebibliography}[1]{\begin{oldthebibliography}{#1}\setlength{\itemsep}{-.9ex}\small}{\end{oldthebibliography}}  
}

\newcommand\esp{{\footnotesize\textsf{\upshape ESP}}}
\newcommand\mdl{{\footnotesize\textsf{\upshape MDL}}}

\renewcommand\S{\mathcal S}
\newcommand\C{\mathcal C}

\newcommand\eps{\varepsilon}
\newcommand\st{s.\,t.}
\newcommand\ie{i.\,e.}
\newcommand\eg{e.\,g.}
\newcommand\cf{c.\,f.}
\newcommand\wolog{w.\,l.\,o.\,g.}

\newcommand\rhs{r.\,h.\,s.}
\newcommand\wrt{w.\,r.\,t.}
\newcommand\lref[1]{Lemma~\ref{#1}}
\newcommand\block[1]{\smallskip\par\noindent\makebox{\sffamily\textsl{#1}}}

\usepackage{graphicx}

\begin{document}

%%%%%%%%%%%%%%%%%%%%%%%%%%%%%%%%%%%%%%%%%%%%%%%%%%%%%%%%%%%%%%%%%%%%%%%%%%%%%%%%

%%% Titelblock %%%
\title{\Large{\sffamily\bfseries On Probability Estimation by Exponential Smoothing}}
\author{
	\vspace{.5em}
	Christopher Mattern\\  
  \normalsize\textit{Technische Universit\"at Ilmenau}\\
	\normalsize\textit{Ilmenau, Germany}\\
	\small\texttt{christopher.mattern@tu-ilmenau.de}
}
\date{}	% kein Datum anzeigen

\maketitle
\thispagestyle{empty}

%%% Abstract %%%

\vspace{-30pt}
\begin{center}
\parbox{.95\textwidth}{
\small
\paragraph{Abstract.}
Probability estimation is essential for every statistical data compression algorithm.
In practice probability estimation should be adaptive, \ie\ recent observations should receive a higher weight than older observations.
We present a probability estimation method based on exponential smoothing that satisfies this requirement and runs in constant time per letter.
Our main contribution is a theoretical analysis in case of a binary alphabet for various smoothing rate sequences:
We show that the redundancy \wrt\ a piecewise stationary model with $s$ segments is $O\left(s\sqrt n\right)$ for any bit sequence of length $n$, an improvement over redundancy $O\left(s\sqrt{n\log n}\right)$ of previous approaches with similar time complexity.
}
\end{center}

%%%%%%%%%%%%%%%%%%%%%%%%%%%%%%%%%%%%%%%%%%%%%%%%%%%%%%%%%%%%%%%%%%%%%%%%%%%%%%%%

\section{Introduction}\label{sec:intro}

\paragraph{Background.}
Sequential probability assignment is an elementary component of every statistical data compression algorithm, such as Prediction by Partial Matching, Context Tree Weighting and PAQ (``pack'').
Statistical compression algorithms split compression into modeling and coding and process an input sequence letter-by-letter.
During modeling a model computes a distribution $p$ and during coding an encoder maps the next letter $x$, given $p$, to a codeword of a length close to $-\log p(x)$ bits (this is the ideal code length).
Decoding is the reverse:
Given $p$ and the codeword the decoder restores $x$.
Arithmetic Coding is the de facto standard en-/decoder, it closely approximates the ideal code length \cite{eoit}.
All of the mentioned compression algorithms require simple, elementary models to predict a probability distribution.
Elementary models are typically based on simple closed-form expressions, such as relative letter frequencies.
Nevertheless, elementary models have a big impact on both theoretical guarantees \cite{veness12cts, willems95ctw} and empirical performance \cite{mattern11,veness13ptw} of statistical compression algorithms.
(Commonly, we express theoretical guarantees on a model by the amount of bits the model requires above an ideal competing scheme assuming ideal encoding, the so-called redundancy.)
It is wise to choose elementary models carefully and desirable to analyze them theoretically and to study them experimentally.
In this work we focus on elementary models with the ability to adapt to changing statistics (see next paragraph) whose implementation meets practical requirements, that is $O(Nn)$ (arithmetic) operations and $O(N)$ data words (holding \eg\ a counter or a rational number) while processing a sequence of length $n$ over an alphabet of size $N$.

\paragraph{Previous Work.}
Relative frequency-based elementary models, such as the Laplace- and KT-Estimator, are well-known and well-understood \cite{eoit}. 
A major drawback of these classical techniques is that they don't exploit recency effects (adaptivity):
For an accurate prediction novel observations are of higher importance than past observations \cite{howard91aac,oneil12actw}.
From a theoretical point of view adaptivity is evident in low redundancy \wrt\ an adaptive competing scheme such as a Piecewise Stationary Model (PWS).
A PWS partitions a sequence of length $n$ arbitrarily into $s$ segments and predicts an arbitrary fixed probability distribution for every letter within a segment.
(Since both segmentation and prediction within a segment are arbitrary, we may assume both to be optimal.)

To lift the limitation of classical relative frequency-based elementary models we typically age observation counts, aging takes place immediately before incrementing the count of a novel letter.
For aging frequency-based elementary models there exist two major strategies, which are heavily used in practice.
In \emph{Strategy~1} (count rescaling) we divide all counts by a factor in well-defined intervals (\eg\ when the sum of all counts exceeds a threshold) \cite{howard91aac}, for \emph{Strategy~2} (count smoothing) we multiply all counts by a factor in $(0,1)$ in every update \cite{oneil12actw}.
Strategy~1 was analyzed in \cite{mattern14} and has redundancy $O(s\sqrt n\log n)$.
Similarly, a KT-estimator which completely discards all counts periodically was analyzed in \cite{shamir99} and has redundancy $O(s\sqrt{n\log n})$.
Strategy~2 was studied mainly experimentally \cite{oneil12actw,veness12cts,veness13ptw}.

Another approach for adaptive probability estimation, commonly used in PAQ, is smoothing of probabilities, \emph{Strategy~3}.
Given a probability distribution (\ie\ the prediction of the previous step) and a novel letter we carry out an update as follows:
First we multiply all probabilities with smoothing rate $\alpha\in(0,1)$ and afterwards we increment the probability of the novel letter by $1-\alpha$.
Smoothing rate $\alpha$ does not vary from step to step.
To our knowledge this common-sense approach was first mentioned in \cite{howard91piac}.
A finite state machine that approximates smoothing was analyzed in \cite{meron04fsp} and has redundancy $O(n K^{-2/3})$ \wrt\ PWS with $s=1$, where $K$ is the number of states.

All aforementioned approaches meet practical demands, they require $O(Nn)$ (arithmetic) operations and $O(N)$ data words.
More complex (but unpractical) methods are based on mixtures over elementary models associated to so-called transition diagrams \cite{shamir99, willems96pws} or associated to (PWS-)partitions \cite{veness13ptw}.

\paragraph{Our Contribution.}
In this work we analyze a generalization of strategies~2 and 3 for a binary alphabet.
Based on mild assumptions on sequence $\alpha_1, \alpha_2, \dots$ of smoothing rates ($\alpha_k$ is used for an update after observing the $k$-th letter) we explicitly identify an input sequence with maximum redundancy \wrt\ PWS with $s=1$ and subsequently derive redundancy bounds for $s\geq1$ (Section~\ref{sec:red}).
For PWS with arbitrary $s$ we give redundancy bounds for three choices of smoothing rates in Section~\ref{sec:choice}.
First, we consider a fixed smoothing rate $\alpha = \alpha_1 = \alpha_2 = \dots$ (as in PAQ) and provide $\alpha^*(n)$ that guarantees redundancy $O(s\sqrt n)$ for a sequence of length $n$;
second, we propose a varying smoothing rate, where $\alpha_k\approx\alpha^*(k)$; and
finally a varying smoothing rate that is equivalent to Strategy~2 from the previous section.
By tuning parameters we obtain redundancy $O(s \sqrt n)$ for all smoothing rate choices, an improvement over redundancy guarantees known so far for models requiring $O(N n)$ (arithmetic) operations per input sequence.
Section~\ref{sec:exp} supports our bounds with a small experimental study and finally Section~\ref{sec:end} summarizes and evaluates our results and gives perspectives for future research.

\section{Preliminaries}

\paragraph{Sequences.}
We use $x_{i:j}$ to denote a sequence $x_i x_{i+1} \dots x_j$ of objects (numbers, letters, \dots).
Unless stated differently, sequences are bit sequences (have letters \{0, 1\}).
If $i>j$, then $x_{i:j} := \phi$, where $\phi$ is the empty sequence;
if $j=\infty$, then $x_{i:j} = x_i x_{i+1}\dots$ has infinite length.
For sequence $x_{1:n}$ define $x_{<i} := x_{1:i-1}$ and $x_{\leq i} := x_{1:i}$; we call $x_{1:n}$ \emph{deterministic}, if $x_1=\dots=x_n$, and \emph{non-deterministic}, otherwise.

\paragraph{Code Length and Entropy.}
Code length is measured in bits, thus $\log:=\log_2$.
%For a probability distribution $p$ over $\{0,1\}$, letter $x$ and sequence $x_{1:n}$ we define $\ell(x; p) := -\log p(x)$ and $\ell(x_{1:n}; p) := \ell(x_1; p) + \dots + \ell(x_n; p)$.
For probability distribution $p$ over $\{0,1\}$ and letter $x$ we define $\ell(x; p) := -\log p(x)$.
The binary entropy function is denoted as $H(q) := -q\log(q) - (1-q)\log(1-q)$, for a probability $q$.
For sequence $x_{1:n}$ and relative frequency $q$ of a $1$-bit in $x_{1:n}$ let $h(x_{1:n}) := n H(q)$ be the \emph{empirical entropy} of $x_{1:n}$.

\paragraph{Partitions and Sets.}
Calligraphic letters denote sets.
A partition of a non-empty segment (interval) $(a,b]$ of integers is a set of non-overlapping segments $(i_0, i_1], \dots, (i_{n-1}, i_n]$ \st\ $a=i_0<i_1<\dots<i_n=b$.
The phrase \emph{$k$-th segment} uniquely refers to $(i_{k-1}, i_k]$.

\paragraph{Models and Exponential Smoothing of Probabilities.}
We first characterize the term model from statistical data compression, in order define our modeling method.
A \emph{model} \mdl\ maps a sequence $x_{1:n}$ of length $n\geq0$ to a probability distribution $p$ on $\{0,1\}$.
We define the shorthands $\mdl(x_{\leq n}) := p$ (this is \textbf{not} the probability of sequence $x_{\leq n}$!) and $\mdl(x; x_{\leq n}):=p(x)$.
Model $\mdl$ assigns $\ell(x_{\leq n}; \mdl) := {-}\sum_{1\leq k\leq n} \log \mdl(x_k; x_{<k})$ bits to sequence $x_{\leq n}$.
We are now ready to formally define our model of interest.
\begin{dfn}
For sequence $\alpha_{1:\infty}$, where $0<\alpha_1, \alpha_2, \dots<1$, and probability distribution $p$, where $p(0),p(1)>0$, we define model $\esp = (\alpha_{1:\infty}, p)$ by the sequential probability assignment rule
\(
  \esp(x; x_{\leq k}) =
  \begin{cases}
    \alpha_k \esp(x; x_{<k}) + 1-\alpha_k, & \text{if } k>0 \text{ and } x= x_k\\
    \alpha_k \esp(x; x_{<k}), & \text{if } k>0 \text{ and } x\neq x_k\\
    p(x), & \text{if }k=0    
  \end{cases}
  \text.
  \eqlabel{eq:esp}
\)
\end{dfn}

\noindent
Smoothing rates control the adaption of \esp, large $\alpha_i$'s give high weight to old observations and low weight to new observations, the converse holds for small $\alpha_i$'s.
For our analysis we must assume that the smoothing rates are sufficiently large:

\begin{asm}\label{asm:esp}
$\esp = (\alpha_{1:\infty}, p)$ satisfies $\frac12<\alpha_1, \alpha_2, \dots<1$ and \wolog\ $p(0)\leq p(1)$.
\end{asm}

\noindent
For the upcoming analysis the product of smoothing rates plays an important role.
Hence, given smoothing rate sequence $\alpha_{1:\infty}$, we define $\beta_0=1$ and $\beta_i := \alpha_1\cdot\ldots\cdot\alpha_i$, for $i>0$.

%%%%%%%%%%%%%%%%%%%%%%%%%%%%%%%%%%%%%%%%%%%%%%%%%%%%%%%%%%%%%%%%%%%%%%%%%%%%%%%%

\section{Redundancy Analysis}\label{sec:red}

\paragraph{First Main Result.}

Now we can state our first main result which compares \esp\ to the code length of an optimal fixed code for $x_{1:n}$, that is the empirical entropy $h(x_{1:n})$.
Before we prove the theorem, we discuss its implications.

\begin{thm}\label{thm:esp-red}
If Assumption~\ref{asm:esp} holds, then we have
\(
  &\hspace{-.5em}\ell(x_{1:n}; \esp) - h(x_{1:n}) \\
  &\leq
  \begin{cases}
    \hfill\sum_{i=0}^{n-1} \log\frac1{1-p(1)\beta_i}, & \text{if } x_{1:n} \text{ is deterministic}\\
    \log \frac1{p(1)\beta_{n-1}} + \sum_{i=0}^{n-2}\log\frac1{1-p(1)\beta_i} - n H\left(\frac 1n\right),& \text{otherwise}
  \end{cases}
  \eqlabel{eq:esp-red}
  \text.
\)
\end{thm}

\noindent
Recall that by Assumption~\ref{asm:esp} we have $p(0)\leq p(1)$.
First, consider a deterministic sequence of $0$-bits.
By \eqref{eq:esp} we have $\esp(1; x_{\leq i}) = \beta_i p(1)$, thus $\esp(0; x_{\leq i}) = 1-\beta_i p(1)$, both for $0\leq i<n$.
The total code length is $\ell(x_{1:n}; \esp) = -\sum_{i=0}^{n-1} \log(1-\beta_i p(1))$ and clearly $h(x_{1:n})=0$, so the redundancy $\ell(x_{1:n}; \esp) - h(x_{1:n})$ matches \eqref{eq:esp-red}.
Now consider a non-deterministic sequence $x_{1:n}=00\dots01$ with single $1$-bit at position $n$.
Similar to the deterministic case the same equations for $\esp(\,\cdot\,; x_{\leq i})$ hold, for $i<n$.
The total code length is $\ell(x_{1:n}; \esp) = -\sum_{i=0}^{n-2} \log(1-\beta_i p(1)) -\log(\beta_{n-1} p(1))$ and the empirical entropy is $h(x_{1:n}) = n H(\frac1n)$.
Again, the redundancy matches \eqref{eq:esp-red}.
In summary, if $p(0)\leq p(1)$, then $00\dots0$ is a deterministic sequence with maximum redundancy and $00\dots01$ is a non-deterministic sequence with maximum redundancy.
Similar statements hold, if $p(0)\geq p(1)$: by symmetry we must toggle $0$-bits and $1$-bits.
When $p(0)=p(1)$ we have equal redundancy, \eg\ $\ell(00\dots0;\esp)=\ell(11\dots1; \esp)$ in the deterministic case.
%By Theorem~\ref{thm:esp-red} we can exactly depict the worst-case input $x_{1:n}$ for a given instance of \esp\ by comparing the redundancy of just $2$ sequences.
In summary, for a given instance of \esp\ (that satisfies Assumption~\ref{asm:esp}) the worst-case input is either $00\dots0$ (only $0$-bits) or $00\dots01$ (single $1$-bit), among all $2^n$ bit sequences of length $n$.
For fixed $n$ we can now easily compare the redundancies of those two inputs and immediately depict the worst-case input and its redundancy.

For the proof of Theorem~\ref{thm:esp-red} we require the following lemma.

% \begin{lem}\label{lem:esp-pmin}
% Let $p'$ be a distribution, \st\ $p'(0)\leq p'(1)$ and that there exists $\esp=(\alpha_{1:\infty}, p)$ and a bit $x$ so that we may write $p'=\esp(x)$.
% If $\alpha_1>\frac12$, then $x=0$ minimizes $p(x)$ and $p(x)\leq\frac12$.
% \end{lem}
% \begin{prf}
% We distinguish 3~cases and show that $p(x)$ is minimal and $x=0$ in Case~1.
% \par\block{Case~1: $1-\alpha_1< p'(0) \leq \frac12$ and $x=0$.}
% We have $p(x) = [p'(0)-(1-\alpha_1)]/\alpha_1 =: q_1$.
% \par\block{Case~2: $1-\alpha_1< p'(0) \leq \frac12$ and $x=1$.}
% We have $p(x) = 1-p(0) = [\alpha_1-p'(0)]/\alpha_1 =: q_2$.
% \par\block{Case~3: $0<p'(0)\leq1-\alpha_1$.}
% We must have $x=1$ (if $x=0$, then $p'(0)>1-\alpha_1$) and $p(x) = [\alpha_1-p'(0)]/\alpha_1 \geq [2\alpha_1-1]/\alpha_1=:q_3$.
% \smallskip\par\noindent
% From $p'(0)\leq\frac12$ we conclude $q_1\leq q_2$ and $q_1\leq[\alpha_1-1/2]/\alpha_1$ and from $\alpha_1>\frac12$ we conclude $[\alpha_1-1/2]/{\alpha_1}\leq q_3$, thus $q_1\leq q_3$.
% Since $q_1\leq\frac12$ we have $p(x)\leq\frac12$ in Case~1.
% \end{prf}

\begin{lem}\label{lem:esp-hdiff}
Any non-deterministic sequence $x_{1:n}$ of length $n\geq2$ satisfies
\(
h(x_{1:n})-h(x_{2:n})
\geq
\begin{cases}
  n H\left(\frac1n\right),& \hspace{-.5em}\text{if } x_{2:n} \text{ is deterministic} \\
  n H\left(\frac1n\right)-(n-1) H\left(\frac1{n-1}\right),& \hspace{-.5em}\text{otherwise}
\end{cases}
\text.
\)
\end{lem}
\begin{prf}
Let $1-p$ be the relative frequency of $x_1$ in $x_{1:n}$, thus $h(x_{1:n})-h(x_{2:n}) = n H(p) - (n-1) H\left(\textstyle \frac n{n-1}\cdot p\right) =: f(p)$.
We distinguish two cases:
\par\block{Case~1: $x_{2:n}$ is deterministic.}
We have $p=\frac{n-1}n$ and $f(p)= n H\left(\frac{n-1}n\right) = n H\left(\frac1n\right)$.
\par\block{Case~2: $x_{2:n}$ is non-deterministic.}
Since $H(p)$ is concave, $H'(p)$ is decreasing and $f'(p) = n\left[H'(p)-H'\left(\frac n{n-1}\cdot p\right)\right]\geq0$, \ie\ $f(p)$ is increasing and minimal for minimum $p$.
Since $x_{1:n}$ is non-deterministic the minimum value of $p$ is $\frac1n$ and we get $f(p) \geq f\left(\frac1n\right) = n H\left(\frac1n\right)-(n-1) H\left(\frac1{n-1}\right)$.
\end{prf}

\noindent
Now let us proceed with the major piece of work in this section.

\begin{prf}[\bfseries\sffamily Proof of Theorem~\ref{thm:esp-red}]
We define $r(x_{1:n}, \esp) := \ell(x_{1:n}; \esp)-h(x_{1:n})$ and distinguish:

\block{Case 1: $x_{1:n}$ is deterministic.}
By $p(0)\leq p(1)$ (Assumption~\ref{asm:esp}) we have $\esp(x; x_{<i}) \geq \esp(0; x_{<i}) = 1-p(1)\beta_{i-1}$ and $h(x_{1:n})=0$, we get
\(
  r(x_{1:n}, \esp) = \sum_{1\leq i\leq n}\log\frac1{\esp(x_i; x_{<i})} \leq \sum_{0\leq i<n} \log\frac1{1-p(1)\beta_i}
  \text.
\)

\block{Case 2: $x_{1:n}$ is non-deterministic.}
We have $n\geq2$ and by induction on $n$ we prove
\(
  r(x_{1:n}, \esp) \leq \log \frac1{p(1)\beta_{n-1}} + \sum_{0\leq i<n-1}\log\frac1{1-p(1)\beta_i} - n H\left(\textstyle \frac1n\right)
  \text.
\)

\block{Base: $n=2$. ---}
We have $x_{1:n}\in\{01, 10\}$, in either case $h(x_{1:n})= nH\left(\frac1n\right)=2$ and $\ell(x_{1:n}; \esp) = \log\frac1{p(x_1) \beta_1 p(x_2)} = \log\frac1{p(1)\beta_1} + \log\frac1{1-p(1)}$, the claim follows.
\block{Step: $n>2$. ---}
By defining $\esp' = (\alpha_{1:\infty}', p')$, where $\alpha_i' = \alpha_{i+1}$, $\beta_i' = \alpha_1' \cdot\ldots\cdot\alpha_i'$, $p'=\esp(x_{\leq 1})$ we may write
\(
  r(x_{1:n}, \esp) = \log\frac1{p(x_1)} + r(x_{2:n}, \esp') - (h(x_{1:n})-h(x_{2:n}))
  \text.
  \eqlabel{eq:esp-red-1}
\)
Now \wolog\ fix $p'$ \st\ $p'(0)\leq p'(1)$.
Since we want to bound \eqref{eq:esp-red-1} from above, we must choose $x_1$ \st\ $p(x_1)$ is minimal (and the \rhs\ of \eqref{eq:esp-red-1} is maximal).
To do so, distinguish:
\block{Case~1: $x_1=0$.}
For some distribution $q$ with $q(0)>0$ we have $p(x_1)=q(0)$ and $\frac12 \geq p'(0)=\alpha_1 q(0) + 1-\alpha_1$, thus $q(0)\leq\left[\alpha_1-\frac12\right]/\alpha_1$.
(Notice the subtle detail: $\alpha_1\leq\frac12$ implies $q(0)\leq0$, which contradicts $q(0)>0$ and would make Case~1 impossible; however we assumed $\alpha_1>\frac12$.)
Furthermore, we have $q(0)\leq\frac12$.
\block{Case~2: $x_1=1$.}
For some distribution $r$ with $r(1)>0$ we have $p(x_1)=r(1)$ and $\frac12 \leq p'(1)=\alpha_1 r(1) + 1-\alpha_1$, thus $r(1)\geq\left[\alpha_1-\frac12\right]/\alpha_1$.

Since $q(0)\leq r(1)$ (\ie\ Case~1 minimizes $p(x_1)$) and $q(0)\leq\frac12$ we may now \wolog\ assume that $x_1=0$, $p'(1) = \alpha_1 p(1)$, $p(x_1)=1-p(1)$ and $p(0)\leq p(1)$.
We distinguish:

\block{Case 1: $x_{2:n}$ is deterministic.}
We must have $x_{2:n}=11\dots1$, since $x_1=0$ and $x_{1:n}$ is non-deterministic, thus
\(
  r(x_{2:n}, \esp')
  = \sum_{0\leq i<n-1} \log\frac1{1-p'(0)\beta_i'}
  \leq \log\frac1{p(1)\beta_{n-1}} + \sum_{1\leq i<n-1}\log\frac1{1-p(1)\beta_i}
  \text,
  \eqlabel{eq:esp-red-3}
\)
where we obtain the inequality by $p'(0)\beta_i'\leq p'(1)\beta_i'=p(1)\beta_{i+1}$, for $i<n-2$ and $1-p'(0)\beta_{n-2}'=1-[1-p(1)\alpha_1]\beta_{n-1}/\alpha_1 \geq p(1)\beta_{n-1}$, for $i=n-2$.
To obtain the claim we plug the inequalities \eqref{eq:esp-red-3} and $h(x_{1:n})-h(x_{2:n})\geq n H(\frac1n)$ (by \lref{lem:esp-hdiff}) into \eqref{eq:esp-red-1} and note that $p(x_1) = 1- p(1)\beta_0$ (since $\beta_0=1$).

\block{Case 2: $x_{2:n}$ is non-deterministic.}
The hypothesis and $p'(1)\beta_i' = p(1)\beta_{i+1}$ yield
\(
  r(x_{2:n}, \esp')
  &\leq \log\frac1{p'(1)\beta_{n-2}'} + \hspace{-.2em}\sum_{0\leq i<n-2}\hspace{-.2em}\log\frac1{1-p'(1)\beta_i'}-(n-1)H\left(\textstyle\frac1{n-1}\right)\\
  &= \log\frac1{p (1)\beta_{n-1} } + \hspace{-.2em}\sum_{1\leq i<n-1}\hspace{-.2em}\log\frac1{1-p(1)\beta_{i}}-(n-1)H\left(\textstyle\frac1{n-1}\right)
  \eqlabel{eq:esp-red-4}
  \text.
\)
We plug the inequalities \eqref{eq:esp-red-4} and $h(x_{1:n})-h(x_{2:n}) \geq n H(\frac1n) - (n-1) H(\frac1{n-1})$ (by \lref{lem:esp-hdiff}), into \eqref{eq:esp-red-1} and note that $p(x_1) = 1- p(1)\beta_0$ (since $\beta_0=1$) to end the proof.
\end{prf}

\paragraph{Second Main Result.}
Let us now extend the competing scheme of Theorem~\ref{thm:esp-red}, to which we compare \esp\ to.
Suppose the competing scheme splits the input sequence $x_{1:n}$ according to an arbitrary partition $\S$ of $[1,n]$ and may use an optimal fixed code within every segment $[a,b]\in\S$.
The competing scheme has total coding cost $h(x_{a:b})$ for $x_{a:b}$, thus coding cost $\sum_{[a,b]\in\S} h(x_{a:b})$ for $x_{1:n}$
Notice, that this a lower bound on the coding cost of any PWS with partition $\S$.
Since the situation within a segment resembles the situation of Theorem~\ref{thm:esp-red}, we may now naturally extend the redundancy analysis to the aforementioned competitor.
\begin{thm}\label{thm:esp-red-pws}
Let $\S$ be an arbitrary partition of $[1,n]$.
If Assumption~\ref{asm:esp} holds, then
\(
  \ell(x_{1:n}; \esp) -\sum_{[a,b]\in\S} h(x_{a:b})\leq \lvert S\rvert\log\frac1{p(0)\beta_{n-1}} + \sum_{(a,b]\in\S} \sum_{a<i<b} \log\frac1{1-\beta_i/\beta_a}
  \text.
  \eqlabel{eq:esp-red-pws}
\)
\end{thm}
\begin{prf}%[Proof of Theorem~\ref{thm:esp-red-pws}]
Let $r(x_{1:n}, \esp) := \ell(x_{1:n}; \esp)-h(x_{1:n})$.
Our plan for the proof is to simplify \eqref{eq:esp-red} (see calculations below) to yield
\(
  \ell(x_{1:n}; \esp)-h(x_{1:n})
  \leq \log\frac1{p(0)\beta_{n-1}} + \sum_{1\leq i<n}\log\frac1{1-\beta_i}
  \eqlabel{eq:esp-red-pws-1}
\)
and use to \eqref{eq:esp-red-pws-1} to bound the redundancy for an arbitrary segment $(a,b]$ from $\S$ (see calculations below) via
\(
  \sum_{a<i\leq b} \log\frac1{\esp(x_i; x_{<i})}-h(x_{a+1:b})
  \leq \log\frac1{p(0)\beta_{b-1}} + \sum_{a<i<b}\log\frac1{1-\beta_i/\beta_a}
  \text.
  \eqlabel{eq:esp-red-pws-2}
\)
We now obtain \eqref{eq:esp-red-pws} easily by summing \eqref{eq:esp-red-pws-2} over all segments $(a,b]$ from $\S$ and by $\beta_{b-1}\geq \beta_{n-1}$.
\block{Simplifying \eqref{eq:esp-red}.}
Observe that $\sum_{0\leq i<n}\log\frac1{1-p(1)\beta_i} = \log\frac1{p(0)}+\sum_{1\leq i<n}\log\frac1{1-p(1)\beta_i}$ and furthermore $p(0)\leq p(1)$.
So bound \eqref{eq:esp-red} becomes
\(
  r(x_{1:n}, \esp)
  &\leq\log\frac1{p(0)}+\hspace{-.2em}\sum_{1\leq i<n}\hspace{-.1em}\log\frac1{1-p(1)\beta_i}
  \leq\log\frac1{p(0)} + \hspace{-.2em}\sum_{1\leq i<n}\hspace{-.1em}\log\frac1{1-\beta_i}
  \text,
\intertext{if $x_{1:n}$ is deterministic and by $\log\frac1{p(1)}-nH\left(\frac1n\right)\leq0$ (since $p(1)\geq\frac12$ and $n\geq2$)}
  r(x_{1:n}, \esp)
  &\leq\log\frac1{p(0) p(1) \beta_{n-1}}+\sum_{1\leq i<n-1}\log\frac1{1-p(1)\beta_i}-nH\left(\textstyle \frac1n\right)\\
  &\leq\log\frac1{p(0)\beta_{n-1}} + \sum_{1\leq i<n}\log\frac1{1-\beta_i}
  \text,
\)
if $x_{1:n}$ is non-deterministic.
In either case bound \eqref{eq:esp-red-pws-1} holds.

\block{Redundancy of $(a,b]$.}
For segment $(a,b]$ we define sequence $x'_{1:b-a} = x_{a+1:b}$ and $\esp' = (\alpha_{1:\infty}', p')$, \st\ $\esp(x; x_{<i}) = \esp(x'; x_{<i-a}')$ for $i\in(a,b]$.
Therefore, let $\alpha'_{1:\infty} = \alpha_{a+1:\infty}$, $\beta_i' = \alpha_1'\cdot\ldots\cdot\alpha_i'$ and \wolog\ $p'(0)\leq p'(1)$.
We obtain
\(  
  &\sum_{a<i\leq b} \log\frac1{\esp(x_i; x_{<i})}-h(x_{a+1:b})\\
  &\quad= \sum_{1\leq i\leq b-a}\log\frac1{\esp'(x_i'; x_{<i}')}-h(x_{1:b-a}')
  = \ell(x_{1:n}, \esp)-h(x_{1:n})\\
  &\quad\stackrel{\eqref{eq:esp-red-pws-1}}\leq \log\frac1{p'(0)\beta_{b-a-1}'} + \hspace{-.5em}\sum_{1\leq i<b-a}\hspace{-.5em}\log\frac1{1-\beta_i'}
  \leq \log\frac1{p(0) \beta_{b-1}} + \sum_{a<i<b}\log\frac1{1-\beta_i/\beta_a}
  \text,
\)
where the last step is due to $p'(0)\geq p(0)\beta_a$ (also $p'(1)\geq p(0)\beta_a$) and $\beta_i' = \beta_{a+i}/\beta_a$.
\end{prf}

%%%%%%%%%%%%%%%%%%%%%%%%%%%%%%%%%%%%%%%%%%%%%%%%%%%%%%%%%%%%%%%%%%%%%%%%%%%%%%%%

\section{Choice of Smoothing Rate Sequence}\label{sec:choice}

\paragraph{Fixed Smoothing Rate.}
A straight-forward choice for the smoothing rates is to use the same rate $\alpha$ in every step.
This leads to a simple and fast implementation, since no smoothing rate sequence needs to be computed or stored.
We require the following lemma for the analysis:
\begin{lem}\label{lem:esp-logsum}
For $0<\alpha<1$ we have $\sum_{1\leq i\leq m} \log\frac1{1-\alpha^i} \leq \frac{(\pi \log e)^2}{6\log\frac1\alpha}$.
\end{lem}
\begin{prf}
For $m=0$ the bound trivially holds, let $m\geq1$.
Since $\log\frac1{1-\alpha^z}$ is decreasing in $z$ and integrable for $z$ in $[0,\infty)$ we may bound the series by an integral,
\(
  \sum_{1\leq i\leq m} \log\frac1{1-\alpha^i}
  \leq \int_0^{m} \log\frac1{1-\alpha^z} \mathrm dz
  =  \log(e) \int_0^{m} \sum_{j\geq1} \frac{\alpha^{jz}}j \mathrm dz
  \text.
  \eqlabel{eq:esp-logsum-1}
\)
The equality in \eqref{eq:esp-logsum-1} follows from the series expansion $\ln\frac1{1-y}=\sum_{j\geq1}y^j/j$, for $\lvert y\rvert<1$.
To end the proof, it remains to bound the integral in \eqref{eq:esp-logsum-1} as follows (notice $\sum_{j\geq1} j^{-2} = \pi^2/6$):
\(
  \int_0^{m} \sum_{j\geq1} \frac{\alpha^{jz}}j \mathrm dz
  =\sum_{j\geq1} \frac1j\int_0^{m} \alpha^{jz} \mathrm dz
  =\frac{\log e}{\log\frac1\alpha}\sum_{j\geq1}\frac{1-\alpha^{jm}}{j^2}
  \leq\frac{\pi^2 \log e}{6\log\frac1\alpha}
  \text{.\qedhere}
  \nonumber
\)
\end{prf}

% Bemerkung: Alle Schrittweiten unten erfüllen alpha_1>1/2 (spätestens für genügend große n)
\begin{cor}\label{cor:esp-red-pws-alpha-fixed}
Let $\S$ be an arbitrary partition of $[1,n]$.
If $\alpha=\alpha_1=\alpha_2=\dots$ and Assumption~\ref{asm:esp} holds, then
\(
  \ell(x_{1:n}; \esp)-\sum_{[a,b]\in\S} h(x_{a:b})
  \leq \lvert\S\rvert\cdot\left[\log\frac1{p(0)}+\frac{(\pi\log e)^2}{6\log\frac1\alpha} + (n-1)\log\frac1\alpha\right]
  \text.
  \eqlabel{eq:esp-red-pws-alpha-fixed}
\)
\end{cor}
\begin{prf}
We have $\beta_i=\alpha^i$, thus for $i\in(a,b]$ we plug the estimate
\(
  \sum_{a<i<b}\log\frac1{1-\beta_i/\beta_a}
  = \sum_{0<i-a<b-a}\log\frac1{1-\alpha^{i-a}}
  \hspace{-.5em}\stackrel{\text{Lem. \ref{lem:esp-logsum}}}
  \leq\hspace{-.2em} \frac{(\pi\log e)^2}{6\log\frac1\alpha}
\)
and $\log\beta_{n-1}=(n-1)\log\alpha$ into \eqref{eq:esp-red-pws} to conclude the proof.
\end{prf}

\noindent
Choosing $\alpha = e^{-\frac{\pi}{\sqrt{6(n-1)}}}$ minimizes the \rhs\ of bound \eqref{eq:esp-red-pws-alpha-fixed} and satisfies $\alpha>\frac12$ (Assumption~\ref{asm:esp}), when $n\geq5$.
The optimal choice gives redundancy at most
\(
  \lvert\S\rvert \cdot\left[ \frac{2\pi\log e}{\sqrt 6}\cdot \sqrt n + \log\frac1{p(0)} \right]
  < \lvert\S\rvert \cdot\left[ 3.701 \cdot \sqrt n + \log\frac1{p(0)} \right]
  \text.
  \eqlabel{eq:esp-red-pws-alpha-opt}
\)

\paragraph{Varying Smoothing Rate.}
It is impossible to choose an optimal fixed smoothing rate, when $n$ is unknown.
A standard technique to handle this situation is the doubling trick, which will increase the $\sqrt n$-term in \eqref{eq:esp-red-pws-alpha-opt} by a factor of ${\sqrt 2}/({\sqrt 2-1})\approx3.41$.
However, we can do better by slowly increasing the smoothing rate step-by-step, which only leads to a factor $\sqrt 2\approx1.41$.

\begin{cor}\label{cor:esp-red-pws-alphak}
Let $\S$ be an arbitrary partition of $[1,n]$.
If $\alpha_k = e^{-\pi/{\sqrt{12(k+1)}}}$ (\ie\ $\alpha_k>\frac12$) and Assumption~\ref{asm:esp} holds, then
\(
  \ell(x_{1:n}; \esp)-\sum_{[a,b]\in\S} h(x_{a:b})
  \leq \lvert\S\rvert\cdot\left[\log\frac1{p(0)}+\frac{2\pi\log e}{\sqrt3}\cdot \sqrt n\right]
  \text.
  \eqlabel{eq:esp-red-pws-alphak}
\)
\end{cor}
\begin{prf}
We have $\beta_i = \mathrm{exp}\left(-\frac\pi{\sqrt{12}}\sum_{1<k\leq i+1} k^{-1/2}\right)$ and bound the terms depending on the $\beta_i$'s in \eqref{eq:esp-red-pws} from above.
First, observe that
\(
  \sum_{1<k\leq n} k^{-1/2}
  \leq \int_1^n \frac{\mathrm dz}{\sqrt z}\leq 2\sqrt n
  \stackrel{\text{Def.} \beta_i} \implies \log\frac1{\beta_{n-1}}\leq \frac{\pi\log e}{\sqrt3}\sqrt n
  \text,
  \eqlabel{eq:esp-red-pws-alphak-1}
\)
second, for $a<i<b$ we have $\beta_i/\beta_a = \alpha_{a+1}\cdot \ldots\cdot\alpha_i\leq (\alpha_{n-1})^{i-a}$, since $i<n$ and $\alpha_1, \alpha_2, \dots$ is increasing, consequently we obtain
\(
  \sum_{a<i<b}\log\frac1{1-\beta_i/\beta_a}
  \leq\sum_{a<i<b}\log\frac1{1-(\alpha_{n-1})^{i-a}}
  \hspace{-.6em}\stackrel{\text{Lem. \ref{lem:esp-logsum}}}\leq\hspace{-.3em} \frac{(\pi\log e)^2}{6\log\frac1{\alpha_{n-1}}}
  = \frac{\pi\log e}{\sqrt3}\sqrt n
  \text.
  \eqlabel{eq:esp-red-pws-alphak-2}
\)
We plug \eqref{eq:esp-red-pws-alphak-1} and \eqref{eq:esp-red-pws-alphak-2} into \eqref{eq:esp-red-pws}, the result is \eqref{eq:esp-red-pws-alphak}. 
\end{prf}

\paragraph{Count Smoothing.}
Consider aging Strategy~2 from Section~\ref{sec:intro} with smoothing rate $\lambda\in(0,1)$.
We will now show that Strategy~2 is an instance of \esp.
For $s_0, s_1>0$ we define the smoothed count $s(x; x_{\leq k})$ of bit $x$ and the smoothed total count $t_k$ as follows
\(
  s(x; x_{\leq k}) :=
  \begin{cases}
    \lambda s(x; x_{<k})+1, &\text{if } k>0 \text{ and } x_k=x\\
    \lambda s(x; x_{<k}), &\text{if }  k>0 \text{ and } x_k\neq x\\
    s_x\text, &\text{if } k=0
  \end{cases}
  \text{ and }
  t_k :=
  \begin{cases}
    \lambda t_{k-1} + 1\text, &\text{if } k>0\\
    s_0+s_1\text, &\text{if }k=0
  \end{cases}  
  \text.
\)
Strategy~2 predicts $p(x; x_{\leq k}) = s(x; x_{\leq k})/t_k$.
In case $x_k=x$ we get
\(
  p(x; x_{\leq k})
  = \frac{\lambda s(x; x_{<k})+1}{t_k}
  = \frac{\lambda t_{k-1}}{t_k} \frac{s(x; x_{<k})}{t_{k-1}} + \frac1{t_k} = \frac{t_k-1}{t_k} p(x; x_{<k}) + \frac1{t_k}
  \text,
\)
similarly $p(x; x_{\leq k}) = \frac{t_k-1}{t_k} p(x; x_{<k})$, if $x_k\neq x$.
If we now choose $\alpha_k=\frac{t_k-1}{t_k}$ and $p(x) = \frac{s_x}{s_0+s_1}$ the above sequential probability assignment rule resembles \eqref{eq:esp}.
This insight allows us to adopt our analysis method.
To do so, we require the following technical statement first.

\begin{lem}\label{lem:esp-geo-frac}
For $1\leq a\leq b$ and $0<\lambda<1$ we have $\frac{1-\lambda^a}{1-\lambda^b}\geq \frac ab$.
\end{lem}
\begin{prf}
Let $f(z) := \ln((1-\lambda^z)/z)$, it suffices to prove that $f(a)\geq f(b)$.
By $\ln\lambda^z\geq 1-1/\lambda^z$ we get $f'(z) = \left[(1-\ln\lambda^z)\cdot\lambda^z-1\right]/\left[a(1-\lambda^a)\right]\leq0$, so $f$ is decreasing.
\end{prf}

\begin{cor}\label{cor:esp-red-pws-esf}
Let $\S$ be an arbitrary partition of $[1,n]$.
Fix $0<\lambda<1$ and $m\geq1$, define $t_k:= \lambda t_{k-1}+1$ for $k\geq1$ and $t_0=1+\lambda+\dots+\lambda^{m-1}$ for $k=0$.
If $\alpha_k = \frac{t_k-1}{t_k}$ and Assumption~\ref{asm:esp} holds, then
\(
  \ell(x_{1:n}; \esp)
  -\sum_{[a,b]\in\S} h(x_{a:b})
  \leq \lvert\S\rvert\cdot\left[\log\frac n{p(0)}+\frac{(\pi\log e)^2}{6\log\frac1\lambda} + (n-1)\log\frac1\lambda\right]
  \text.
  \eqlabel{eq:esp-red-pws-esf}
\)
\end{cor}
\begin{prf}
Let $k\geq1$ and note that by $t_k = \lambda t_{k-1}+1$ we may write $\alpha_k = \lambda t_{k-1}/t_k$ and $t_k = 1 + \lambda + \dots + \lambda^{k+m-1} = (1-\lambda^{k+m})/({1-\lambda})$ and get
\(
  \beta_i
  = \alpha_1\cdot\ldots\cdot\alpha_i
  =\frac{\lambda t_0}{t_1} \frac{\lambda t_1}{t_2} \dots \frac{\lambda t_{i-1}}{t_i}
  = \frac{t_0}{t_i}\cdot\lambda^i
  = \frac{1-\lambda^m}{1-\lambda^{m+i}}\cdot\lambda^i
  \text.
\)
We now proceed by bounding the terms dependent on $\beta_i$ in \eqref{eq:esp-red-pws}:
\(
  \beta_i
  = \frac{(1-\lambda^m)\lambda^i}{1-\lambda^{m+i}}
  {\stackrel{\text{Lem. \ref{lem:esp-geo-frac}}}\geq} \hspace{-.1em} \frac{m\lambda^i}{m+i}
  \stackrel{m\geq1}\geq \frac{\lambda^i}{i+1}
  \hspace{.5em}\text{and}\hspace{.5em}
 \frac{\beta_i}{\beta_a}
   = \frac{1-\lambda^{m+a}}{1-\lambda^{m+i}}\cdot \lambda^{i-a}  
   \stackrel{a\leq i}\leq \lambda^{i-a}
\)
From the above inequalities we obtain
\(
  \log\frac1{\beta_{n-1}}
  \leq \log\frac n{\lambda^{n-1}}
  \hspace{.5em}\text{and}\hspace{.2em}
  \sum_{a<i<b}\hspace{-.1em} \log\frac1{1-\beta_i/\beta_a}
  \leq \hspace{-.3em}\sum_{a<i<b}\hspace{-.1em}\log\frac1{1-\lambda^{i-a}}
  \hspace{-.4em}{\stackrel{\text{Lem. \ref{lem:esp-logsum}}}\leq} \frac{(\pi\log e)^2}{6\log\frac1\lambda}
  \text.
\)
Finally we plug the above inequalities into \eqref{eq:esp-red-pws} and rearrenging yields \eqref{eq:esp-red-pws-esf}.
\end{prf}

\noindent
For $k\rightarrow\infty$ we have $t_k\rightarrow \frac1{1-\lambda}$, thus $\alpha_k \rightarrow\lambda$, \ie\ we expect the smoothed counts method to perform similar to \esp\ with fixed smoothing rate $\lambda$, when the input is large enough.
Bound \eqref{eq:esp-red-pws-esf} reflects this behavior, it differs from \eqref{eq:esp-red-pws-alpha-fixed} only by the additive term $\lvert\S\rvert\log n$.
Furthermore, the optimal value of $\lambda$ in \eqref{eq:esp-red-pws-esf} matches the optimal value of $\alpha$ in \eqref{eq:esp-red-pws-alpha-fixed}.

%%%%%%%%%%%%%%%%%%%%%%%%%%%%%%%%%%%%%%%%%%%%%%%%%%%%%%%%%%%%%%%%%%%%%%%%%%%%%%%%

\section{Experiments}\label{sec:exp}

For inputs of length $n$ we experimentally checked the tightness of our bounds from the previous section for a wide range of \esp-instances with smoothing rate choices
(i) fixed ``optimal'' smoothing rate $\alpha=\mathrm{exp}(-\pi/\sqrt{6(n-1)})$ (here ``optimal'' means that the corresponding bound, \cf\ Corollary~\ref{cor:esp-red-pws-alpha-fixed}, is minimized),
(ii) varying smoothing from Corollary~\ref{cor:esp-red-pws-alphak} and
(iii) varying smoothing from Corollary~\ref{cor:esp-red-pws-esf} with ``optimal'' $\lambda=\mathrm{exp}(-\pi/\sqrt{6(n-1)})$ and $m=1$.
Since our bounds from corollaries \ref{cor:esp-red-pws-alpha-fixed}, \ref{cor:esp-red-pws-alphak} and \ref{cor:esp-red-pws-esf} are worst-case bounds we compare them to the empirically measured (approximate) worst-case redundancy.
Furthermore, we compare the (approximate) worst-case redundancy of (i), (ii) and (iii) to each other.
We now explain the details below.

\begin{figure}
  \centering
  \includegraphics[width=\linewidth]{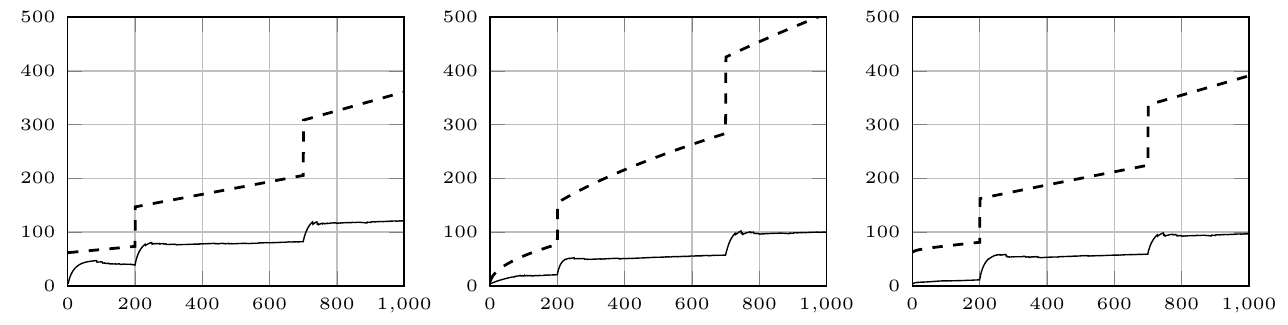}  
  \parbox{.95\textwidth}{
  \renewcommand\figurename{\small\sffamily\bfseries Figure}
  \caption{  
  \textnormal{
  Redundancy bound (dashed line) and approximate worst-case redundancy $r(k)$ (solid line) of class 
  $\{(\alpha_{1:\infty}, p)\mid 0<\eps\leq p(0),p(1)\}$ for $\eps=0.05$  
  \wrt\ competitor with partition $\S=\{[1,200], (200,700], (700,1000]\}$ on the length-$k$ prefix, $1\leq k\leq n$ of a sequence with length $n=1000$.
  The $x$-axis is prefix length $k$ and the $y$-axis is redundancy in bit.
  Every plot corresponds to a different smoothing rate choice:
  (i) fixed ``optimal'' smoothing rate $\alpha=\mathrm{exp}(-\pi/\sqrt{6(n-1)})$,
  (ii) varying smoothing from Corollary~\ref{cor:esp-red-pws-alphak} and
  (iii) varying smoothing from Corollary~\ref{cor:esp-red-pws-esf} with ``optimal'' $\lambda=\mathrm{exp}(-\pi/\sqrt{6(n-1)})$ and $m=1$.
  \label{fig:esp-red}
  }}}
\end{figure}

\paragraph{Experimental Setup.}
In the following let smoothing rate sequence $\alpha_{1:\infty}$, input length $n=1000$, partition $\S=\{(0,200], (200,700], (700,1000]\}$ and $\eps=0.05$ be fixed.
(We inspected the outcome of our experiments for different parameters and got similar results, hence these values.)
We want to judge on our bounds on a wide range of \esp-instances, in particular we choose class $\C = \{(\alpha_{1:\infty}, p)\mid 0<\eps\leq p(0),p(1)\}$ of \esp-instances.
To do so, we have to modify our bound slightly, we must replace the term $p(0)$ by $\eps$:
For instance, in Situation (i), we may bound the redundancy of any $\esp\in\C$ of prefix $x_{1:k}$ of given $x_{1:n}$ as follows
\(
  \ell(x_{1:k}; \esp) - \sum_{[a,b]\in\S} h(x_{a:\min\{k,b\}})
  \leq \lvert\S\rvert \cdot\left[ \frac{2\pi\log e}{\sqrt 6}\cdot \sqrt{k-1} + \log\frac1{\eps} \right]
  \text,  
  \eqlabel{eq:esp-class-bound}
\)
for $1\leq k\leq n$.
Since the resulting bounds remain worst-case bounds, we compare the resulting bounds for situations (i)-(iii) to the worst-case redundancy
\(
  r(k) := \max_{\esp\in\C, x_{1:n}} \Big(\ell(x_{1:k}; \esp) - \sum_{[a,b]\in\S} h(x_{a:\min\{k,b\}})\Big)
  \text.
  \eqlabel{eq:esp-exp}
\)
Unfortunately, computing the maximum is intractable, since $\C$ is uncountably infinite and there are exponentially many sequences $x_{1:n}$.
To lift this limitation we take the maximum over a finite subset of \esp-instances from $\C$ and inputs $x_{1:n}$, specified as follows:
For numbers $q_0,\dots,q_{\lvert\S\rvert}\in\{0.05, \dots, 0.95\}$ we consider pairs $(\esp, x_{1:n})$ \st\
$\esp(0;\phi)=q_0$ ($q_0$ determines an \esp-instance) and 
$x_{1:n}$ is drawn uniform at random from all sequences where for the $i$-th segment $[a,b]\in\S$ subsequence $x_{a:b}$ has exactly $\lfloor q_i\cdot(b-a+1) \rfloor$ $1$-bits ($q_i$ determines the (approximate) fraction of $1$-bits in the $i$-th segment).
We now take the maximum in \eqref{eq:esp-exp} over all combinations $(q_0, \dots, q_{\lvert\S\rvert})$ and repeat the random experiment $100$ times for every combination $(q_0, \dots, q_{\lvert\S\rvert})$ (in total $19^{\lvert\S\rvert+1}\cdot 100$ simulations).
Figure~\ref{fig:esp-red} depicts the approximation of $r(k)$ (solid lines) and our bounds on $\ell(x_{\leq k};\esp)-\sum_{[a,b]\in\S} h(x_{a:\min\{b,k\}})$ (dashed lines).
(For instance, bound \eqref{eq:esp-class-bound} is depicted as dashed line in the left plot of Figure~\ref{fig:esp-red}.)

\paragraph{Approximate Worst-Case Redundancy.}
We now compare (approximate) $r(k)$ for smoothing rate choices (i)-(iii) and observe:
On one hand, as long as $k$ is small, varying smoothing rates, (ii) and (iii), yield lower redundancy than (i), and (iii) performs better than (ii).
On the other hand, when $k$ is large (i), (ii) and (iii) don't differ too much.
The increase in redundancy at $k=201$ and $k=701$ is nearly identical in all cases, the difference in redundancy is almost entirely caused by segment $(0,200]$.

\paragraph{Bounds Behavior.}
Now we compare the bounds to (approximate) $r(k)$.
In general, the tightness of our bounds decreases as the number of segments increases.
This is plausible, since we essentially concatenated the worst-case bound for $\lvert\S\rvert=1$.
However, we don't know, whether or not the worst-case redundancy for $\lvert\S\rvert=1$ can appear in multiple adjacent segments at the same time.
Experiments indicate that this may not be the case.
Furthermore, in (i) the bound is tightest, especially within segment $(0,200]$.
In cases (ii) and (iii) the bounds are more loose.
An explanation is, that in the corresponding proofs we worked with rather generous simplifications, \eg\ when bounding $-\sum_{a<i<b}\log(1-\beta_i/\beta_a)$ from above.
If we compare (ii) to (i) and to (iii) we can see, that bound (ii) is tighter for very small $k$.
The reason is simple:
Bound (ii) does not depend on a smoothing rate parameter, whereas (i) contains the term $1/\log\frac1\alpha$ and (iii) contains the term $1/\log\frac1\lambda$.
These terms dominate the bounds, when $k$ is small and $\alpha$ and $\lambda$ are close to $1$.
(We have $\alpha=\lambda\approx0.96$, since $\alpha$ and $\lambda$ were chosen to minimize the corresponding bound for $n=1000$.)

%%%%%%%%%%%%%%%%%%%%%%%%%%%%%%%%%%%%%%%%%%%%%%%%%%%%%%%%%%%%%%%%%%%%%%%%%%%%%%%%

\vspace{-6pt}
\section{Conclusion}\label{sec:end}

In this work we analyzed a class of practical and adaptive elementary models which assign probabilities by exponential smoothing, \esp.
Our analysis is valid for a binary alphabet.
By choosing smoothing rates appropriately our strategy generalizes count smoothing (Strategy~2) and probability smoothing from PAQ (Strategy~3).
Due to its low memory footprint and linear per-sequence time complexity \esp\ is attractive from a practical point of view.
From a theoretic point of view \esp\ is attractive as well:
For various smoothing rate sequences \esp\ has redundancy only $O(s\sqrt n)$ above any PWS with $s$ segments, an improvement over previous approaches.
A short experimental study supports our bounds.

Nevertheless, experiments indicate that there is room for an improved analysis.
Despite minor technical issues a major approach would be to obtain redundancy bounds \wrt\ PWS that take the similarity of adjacent segments into account.
That is, if adjacent segments have very similar distributions, the increase in redundancy should be small, compared to adjacent segments with drastically different distributions.
Furthermore, it is desirable to generalize the analysis to a non-binary alphabet.
We defer a thorough experimental study that compares \esp\ to other adaptive elementary models to future research.

\paragraph{Acknowledgement.}
The author thanks Martin Dietzfelbinger, Martin Aum\"uller and the anonymous reviewers for helpful comments and suggestions that improved this paper.

%%%%%%%%%%%%%%%%%%%%%%%%%%%%%%%%%%%%%%%%%%%%%%%%%%%%%%%%%%%%%%%%%%%%%%%%%%%%%%%%

\bibliographystyle{plain}
\bibliography{ref}{}

\begin{thebibliography}{10}

\bibitem{eoit}
Thomas~M. Cover and Joy~A. Thomas.
\newblock {\em Elements of {I}nformation {T}heory}.
\newblock Wiley-Interscience, 2nd edition, 2006.

\bibitem{howard91aac}
Paul~G. Howard and Jeffrey~S. Vitter.
\newblock Analysis of arithmetic coding for data compression.
\newblock In {\em Proc. Data Compression Conference}, volume~1, pages 3--12,
  1991.

\bibitem{howard91piac}
Paul~G. Howard and Jeffrey~S. Vitter.
\newblock {P}ractical {I}mplementations of {A}rithmetic {C}oding.
\newblock Technical report, Brown University, USA, 1991.

\bibitem{mattern11}
Christopher Mattern.
\newblock {C}ombining {N}on-stationary {P}rediction, {O}ptimization and
  {M}ixing for {D}ata {C}ompression.
\newblock In {\em Proc. International Conference on Data Compression,
  Communications and Processing}, volume~1, pages 29--37, 2011.

\bibitem{mattern14}
Christopher Mattern.
\newblock {O}n {P}robability {E}stimation via {R}elative {F}requencies and
  {D}iscount.
\newblock 2013.
\newblock http://arxiv.org/abs/1311.1723.

\bibitem{meron04fsp}
Eado Meron and Meir Feder.
\newblock Finite-memory {U}niversal {P}rediction of {I}ndividual {S}equences.
\newblock {\em IEEE Trans. on Information Theory}, 50:1506--1523, 2006.

\bibitem{oneil12actw}
Alexander O'Neill, Marcus Hutter, Wen Shao, and Peter Sunehag.
\newblock {A}daptive {C}ontext {T}ree {W}eighting.
\newblock In {\em Proc. Data Compression Conference}, volume~22, pages
  317--326, 2012.

\bibitem{shamir99}
Gil~I. Shamir and Neri Merhav.
\newblock Low-complexity sequential lossless coding for piecewise-stationary
  memoryless sources.
\newblock {\em IEEE Trans. on Information Theory}, 45:1498--1519, 1999.

\bibitem{veness12cts}
Joel Veness, Kee~Siong Ng, Marcus Hutter, and Michael~H. Bowling.
\newblock {C}ontext {T}ree {S}witching.
\newblock In {\em Proc. Data Compression Conference}, volume~22, pages
  327--336, 2012.

\bibitem{veness13ptw}
Joel Veness, Martha White, Michael Bowling, and A.~Andr\'as~Gyorgy.
\newblock Partition tree weighting.
\newblock In {\em Proc. Data Compression Conference}, volume~23, pages
  321--330, 2013.

\bibitem{willems95ctw}
Frans Willems, Yuri~M. Shtarkov, and Tjalling~J. Tjalkens.
\newblock The context-tree weighting method: basic properties.
\newblock {\em IEEE Trans. on Information Theory}, 41:653--664, 1995.

\bibitem{willems96pws}
Frans M.~J. Willems.
\newblock Coding for a binary independent piecewise-identically-distributed
  source.
\newblock {\em IEEE Trans. on Information Theory}, 42:2210--2217, 1996.

\end{thebibliography}

\end{document}